\documentclass[10pt,conference]{IEEEtran}
\usepackage{cite}
\usepackage{amsmath,amssymb,amsfonts}
\usepackage{graphicx}
\usepackage{textcomp}
\usepackage{xcolor}
\usepackage[hyphens]{url}
\usepackage{fancyhdr}
\usepackage{hyperref}
\usepackage{algorithm}
\usepackage{algpseudocode}
\usepackage{booktabs}
\usepackage{multirow}
\usepackage{multicol}
\algnewcommand\algorithmicoutput{\textbf{Output:}}
\algnewcommand\Output{\item[\algorithmicoutput]}
\usepackage{subcaption}
\usepackage{tikz}

\newcommand{\hpcayear}{2027}

\newcommand{\hpcasubmissionnumber}{1560}
\newcommand{\paperName}{YAVIN}
\newcommand{\paperLongName}{A Unified Architecture for Secure Edge Processing in Memory}
\title{\paperName: \paperLongName}
\def\hpcacameraready{} 

\newcommand\hpcaauthors{Shouzhi Fang$\dagger$, William C. Tegge$\dagger$, Md Omar Faruque$\dagger$, Peipei Zhou$\ddagger$, Endadul Hoque$\dagger$, Alex K. Jones$\dagger$}
\newcommand\hpcaaffiliation{Syracuse University$\dagger$, Brown University$\ddagger$}
\newcommand\hpcaemail{
  \{sfang18, wtegge, mdfaruqu, enhoque, akj\}@syr.edu$\dagger$, 
  peipei\_zhou@brown.edu$\ddagger$, 
}

\usepackage{ifthen}
\usepackage{xcolor}
\usepackage{mathabx}
\newboolean{showcomments}
\setboolean{showcomments}{true}

\makeatletter
\newcommand{\mynote}[3]{%
  \ifthenelse{\boolean{showcomments}}{%
   \fbox{\bfseries\sffamily\scriptsize#1}%
   {\small$\blacktriangleright$\textsf{\emph{\color{#3}{#2}}}$\blacktriangleleft$}}%
  {%
   \@bsphack
   \@esphack
  }%
}
\makeatother

\definecolor{asparagus}{rgb}{0.53, 0.66, 0.42}

\author{
  \ifdefined\hpcacameraready
    \IEEEauthorblockN{\hpcaauthors{}}
      \IEEEauthorblockA{
        \hpcaaffiliation{} \\
        \hpcaemail{}
      }
  \else
    \IEEEauthorblockN{\normalsize{HPCA \hpcayear{} Submission
      \textbf{\#\hpcasubmissionnumber{}}} \\
      \IEEEauthorblockA{
        Confidential Draft \\
        Do NOT Distribute!!
      }
    }
  \fi 
}

\fancypagestyle{camerareadyfirstpage}{%
  \fancyhead{}
  
  \fancyhead[C]{
    \ifdefined\aeopen
    \parbox[][12mm][t]{13.5cm}{\hpcayear{} IEEE International Symposium on High-Performance Computer Architecture (HPCA)}    
    \else
      \ifdefined\aereviewed
      \parbox[][12mm][t]{13.5cm}{\hpcayear{} IEEE International Symposium on High-Performance Computer Architecture (HPCA)}
      \else
      \ifdefined\aereproduced
      \parbox[][12mm][t]{13.5cm}{\hpcayear{} IEEE International Symposium on High-Performance Computer Architecture (HPCA)}
      \else
      \parbox[][0mm][t]{13.5cm}{\hpcayear{} IEEE International Symposium on High-Performance Computer Architecture (HPCA)}
    \fi 
    \fi 
    \fi 
    \ifdefined\aeopen 
      \includegraphics[width=12mm,height=12mm]{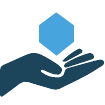}
    \fi 
    \ifdefined\aereviewed
      \includegraphics[width=12mm,height=12mm]{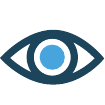}
    \fi 
    \ifdefined\aereproduced
      \includegraphics[width=12mm,height=12mm]{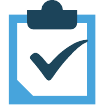}
    \fi
  }
  \fancyfoot[C]{}
}
\begin{document}
\maketitle

\ifdefined\hpcacameraready 
  \thispagestyle{camerareadyfirstpage}
  \pagestyle{empty}
\else
  \thispagestyle{plain}
  \pagestyle{plain}
\fi

\newcommand{\hpcaheight}{0mm}
\ifdefined\eaopen
\renewcommand{\hpcaheight}{12mm}
\fi

\begin{abstract}

Secure, private multi-tenant execution spanning processors, memory, and accelerators remains one of the most significant challenges in modern edge computing systems. Simultaneously, processing-in-memory (PIM) has emerged as an effective approach for reducing the von Neumann bottleneck by moving computation closer to data. Existing trusted execution environments (TEEs) establish trust only within the processor, protecting data while it traverses untrusted resources such as the memory bus. Consequently, trusted computation cannot be performed directly within memory. At the opposite extreme, fully homomorphic encryption enables computation on encrypted data but remains impractical for these multi-tenant systems because of its substantial performance and storage overheads. We present YAVIN, a unified trusted computing base (TCB) that extends the trusted execution environment beyond the processor to encompass both processor execution and a dedicated memory region supporting trusted processing-in-memory execution while treating the memory bus as untrusted. Leveraging the dedicated protected memory regions already established by conventional TEE architectures, YAVIN enables data to be decrypted, processed, and re-encrypted by either processor or PIM execution while remaining within the trusted execution environment, without exposing plaintext on the shared memory bus. To realize this unified TCB, YAVIN presents the first PIM implementations of the LightSaber post-quantum public-key cryptosystem and ASCON-128 authenticated encryption, co-designing both algorithms for efficient DRAM execution to establish and maintain shared cryptographic state, provide authenticated communication over an untrusted memory bus, and support trusted execution spanning the processor and memory. Finally, we demonstrate how cryptography-PIM co-design for tensor-based workloads reorganizes computation to satisfy the ordering constraints imposed by authenticated encryption with minimal performance overhead while simultaneously enabling bit-sliced ordering that limits temporary plaintext exposure. Compared to the latest PIM AES implementation, \paperName{} achieves more than a 20$\times$ speedup while incurring only 34\% and 9.3\% overhead when executing INT8 and INT32 quantized edge-class LLMs, respectively, relative to plaintext execution.

\end{abstract}

\section{Introduction}

Distributed computing infrastructures spanning cloud and edge systems have become the dominant platform for computationally intensive workloads, including machine learning, computer vision, scientific computing, and AI inference. To maximize resource utilization, these systems increasingly rely on multi-tenant execution, where multiple users share processors, accelerators, and memory resources while maintaining logical isolation. Emerging Processing-in-Memory (PIM) architectures further improve efficiency by moving computation closer to data, substantially reducing the performance and energy costs associated with data movement. However, these deployments also introduce significant security challenges. Client data and cryptographic material are processed on infrastructure that may be physically accessible, remotely administered, or shared with untrusted tenants, while operating systems, hypervisors, communication channels, and shared hardware resources remain potential attack surfaces capable of exposing sensitive information through software, microarchitectural, or physical attacks~\cite{islam2017exploiting,shi2011limiting,elnaggar2019multi}.

Protecting sensitive computation in these environments remains a fundamental challenge. Existing approaches generally fall into two categories. Trusted Execution Environments (TEEs), such as Intel SGX\cite{costan2016intel}, AMD SEV\cite{amd-sev-snp}, and ARM CCA\cite{arm-cca}, protect computation by isolating trusted software execution from an untrusted operating system and hypervisor while encrypting memory outside the trusted boundary. Although highly efficient, these approaches fundamentally assume a processor-centric execution model in which computation occurs after complete plaintext operands have been reconstructed within the trusted processor. At the opposite extreme, Fully Homomorphic Encryption (FHE) enables computation directly on encrypted data, eliminating the need to expose plaintext during execution~\cite{gentry2009fully}. However, these strong confidentiality guarantees come at the cost of several orders of magnitude higher computational complexity, substantially increased memory capacity requirements, and high energy consumption. 
These overheads are challenging even in cloud datacenters, let alone on edge computing platforms where compute capability, memory capacity, and energy budgets are inherently constrained.
Consequently, neither conventional processor-centric TEEs nor FHE adequately address the security requirements of emerging PIM architectures, motivating a distributed trusted execution model that spans both processors and memory-resident computation.

In addition to introducing new security challenges, DRAM-based PIM~\cite{seshadri2017ambit,hajinazar2021simdram,oliveira2024mimdram,fcdram,elp2im,manyrowact} fundamentally changes how data and computation are organized to achieve high performance through massive parallelism. PIM implementations rely on bit-sliced data representations, structured data movement, and application-specific execution order. In contrast, the execution model assumed by conventional cryptographic algorithms and processor-centric trusted execution environments (TEEs) differs fundamentally from that of efficient PIM implementations. Furthermore, learning-based inference has emerged as the dominant computational workload for edge platforms, including large language models (LLMs), vision transformers, multimodal AI, and related applications. Efficient execution of these workloads on edge platforms, including PIM architectures, further requires tensor reorganization, data quantization, and carefully orchestrated data movement to minimize memory traffic. Reconciling the conflicting requirements of efficient PIM execution, conventional secure execution, and learning-oriented data organization presents a significant challenge. 
Thus, efficient secure execution on PIM-based edge systems requires the co-design of cryptographic protection, trusted execution, data organization, and learning-oriented computation.

\begin{figure}[tbp]
    \centering
    \includegraphics[width=\linewidth]{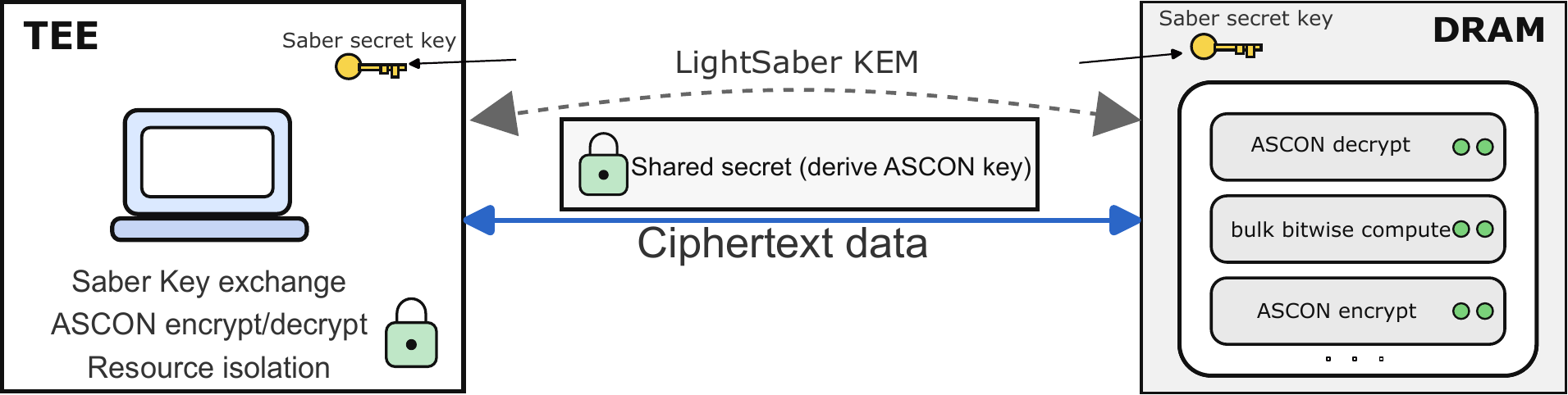}
    \caption{YAVIN unified processor, in-memory, trusted execution environment }
    \label{fig:YAVIN-diagram}
    \vspace{-.1in}
\end{figure}


In this paper we present \paperName, or \paperLongName. \paperName{} creates a distributed TEE that enables cooperative execution between processor- and memory-resident trusted execution environments. The YAVIN concept is illustrated in Fig.~\ref{fig:YAVIN-diagram}. Each tenant of the system can establish a TEE that spans both the processor and the memory system. The memory bus is considered a shared and untrusted resource. Each tenant's TEE is provided with physically isolated trusted memory space as is common within existing TEEs. \paperName{} provides the capability to allow in-memory computation on this area. To establish cooperative operation in the TEE, \paperName{} uses the LightSaber post-quantum key encapsulation mechanism (KEM)~\cite{lightsaber} to establish trusted communication of a shared secret between the processor and the memory system. 

Both the CPU and memory (using PIM execution) generate independent LightSaber public/private key pairs, establishing the trust foundation for the unified trusted computing base (TCB). The memory-side key pair is derived from a DRAM physically unclonable function (PUF)~\cite{puf}, providing a hardware root of trust. The resulting shared secret becomes the symmetric key used to protect communication between the processor and memory within the TCB. \paperName{} further extends the key establishment protocol with a lightweight authentication mechanism that securely supports key initialization and subsequent key updates.

\paperName{} uses the Ascon authenticated encryption with associated data (AEAD) standard~\cite{Ascon} due to its suitability for PIM implementation and its built-in authentication capability. Both the CPU and memory can decrypt data, performs computation on plaintext data, and immediately re-encrypt the results locally, ensuring that plaintext data never traverses the untrusted memory bus. YAVIN provides PIM optimized realizations of both LightSaber and Ascon-AEAD to provide approximately 128-bit security while relying on computational primitives that map efficiently to DRAM-based PIM execution. In addition to the efficiency benefits of Ascon-AEAD over the more standard AES crytography standard and the existing in-memory blockcipher designs~\cite{AESPIM}, Ascon-AEAD provides authentication in addition to cryptography.

As part of its authentication design, Ascon-AEAD requires encryption and decryption to occur in a well-defined order. Rather than treating this as a limitation, \paperName{} exploits the inherent flexibility across two different scheduling dimensions: the order in which tensor indices are evaluated and the order in which bit-level computations are performed. Tensor index scheduling organizes computation into efficient encryption and decryption streams, while bit-level scheduling limits the amount of plaintext simultaneously exposed during execution. For many tensor computations, the order in which independent tensor indices are evaluated may be changed without affecting the final result. \paperName{} exploits this flexibility to organize computation into continuous encryption and decryption streams. 

PIM execution inherently operates on plaintext data. Accordingly, our threat model assumes that the TCB includes a private region of memory in which data may be decrypted, processed, and re-encrypted. Outside this protected execution region, data remains encrypted. However, to further reduce plaintext exposure, \paperName{} exploits the flexibility in execution order, particularly by scheduling bit-level computation. As a result, complete plaintext values are never simultaneously exposed during execution. Instead, only the subset of bit slices required for the current stage of computation is materialized in plaintext, while the remaining bit slices remain encrypted. This reduces the information available from any transient observation of memory contents.

\paperName{} makes the following contributions.
\begin{itemize}
    \item We present \paperName{}, the first unified trusted execution environment that enables cooperative trusted execution between a host processor and charge-sharing DRAM-based processing-in-memory.

    \item We demonstrate the feasibility of secure in-memory execution through detailed implementations of the LightSaber post-quantum key encapsulation mechanism (KEM) and Ascon-AEAD, providing approximately 128-bit security using cryptographic primitives well suited to bulk-bitwise DRAM execution.

    \item We present the first in-DRAM authenticated encryption framework based on an AEAD primitive that integrates confidentiality and authentication tag generation directly within charge-sharing DRAM rather than relying on separate memory-controller, Direct Memory Access, or metadata-based authentication mechanisms.

    \item We demonstrate that co-designing cryptography, data organization, and computation scheduling enables trusted PIM execution while minimizing plaintext exposure and cryptographic overhead.

    \item We provide a detailed evaluation demonstrating the advantages of the proposed co-designed architecture over PIM implementations based on conventional AES- and ECC-based cryptographic schemes.
\end{itemize}

\section{Background and Related Work}
\label{sec:background}

    This section reviews the foundational concepts underlying \paperName{}:
    Trusted execution environments and
    their memory encryption limitations, including lattice-based key encapsulation, the ASCON lightweight cipher, and DRAM-intrinsic entropy sources. We begin with an overview of charge-based in-DRAM PIM.
     
    \subsection{DRAM-Based Processing-In-Memory}
    \label{sec:bg-pim}

    \begin{figure}[tbp]
        \centering
        \includegraphics[width=\linewidth]{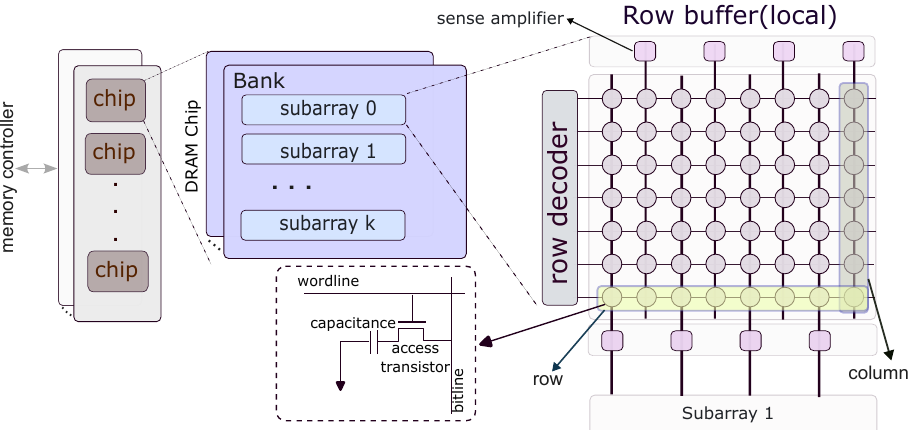}
        \caption{DRAM open-bitline architecture overview}
        \label{fig:dram}
        \vspace{-.1in}
    \end{figure}

    The term PIM is often used to refer to both near-memory computing and processing-using-memory architectures, both with the goal of reducing the data movement bottleneck
    that dominates the energy and latency of memory-intensive
    workloads~\cite{mutlu2019processing, ghose2019processing}.
    \paperName{} targets processing-using memory to perform \emph{bulk-bitwise} PIM, which exploits the
    analog charge-sharing behavior of DRAM bitlines to perform logic
    operations without dedicated computational logic.
     
    In-DRAM PIM primarily targets subarrays within open-bitline memory architectures like the one shown in Fig.~\ref{fig:dram}. A DRAM subarray consists of rows of capacitive cells connected to shared
    bitlines via access transistors. Row buffers are built from sense amplifiers (senseamps) designed to amplify the charge stored in those capacitors. Prior to access, the bitlines are equalized to $\frac{1}{2} V_{DD}$ through a \texttt{PRECHARGE} command.
    An \texttt{ACTIVATE} command turns on the access transistors of a row such that the capacitors perturb the bitline voltage to $\frac{1}{2} V_{DD} \pm \delta$ and are driven to full $V_{DD}$ or $V_{SS}$ logic levels in the senseamps. 
    Bulk-bitwise PIM extends this mechanism with three key primitives:

    \label{sec:bg-ambit}
    \begin{figure}[tbp]
        \centering
        \begin{subfigure}[b]{.66\linewidth}
            \includegraphics[width=\linewidth]{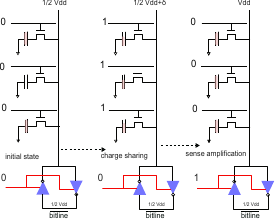}
            \caption{Charge sharing flow}
            \label{fig:charge-sharing}
        \end{subfigure}
        \begin{subfigure}[b]{.32\linewidth}
            \includegraphics[width=\linewidth]{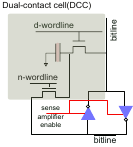}
            \caption{Dual contact cell}
            \label{fig:dcc}
        \end{subfigure}
        \caption{Charge sharing PIM fundamentals}
        \label{fig:charge-sharing-dcc}
    \vspace{-.1in}
    \end{figure}

    \paragraph{RowClone}
    By activating a source row and then a destination row back-to-back
    without an intervening precharge, the charge on the bitlines is
    copied from source to destination~\cite{seshadri2013rowclone}.
    This enables in-DRAM data movement at DRAM-internal bandwidth,
    bypassing the memory bus entirely.
     
    \paragraph{Bulk bitwise logic}
    Simultaneously activating three rows causes their charges to
    compete on the shared bitline as shown in Fig.~\ref{fig:charge-sharing}. In the example, two full capacitors and one empty capacitor still deflect the bitline by $+\delta$. The senseamp drives the bitline to $V_{DD}$ filling all three capacitors, resolving each
    bit to the majority value of the three inputs. This majority operation can be used directly as a majority logic primitive~\cite{MAJ-SYN,hajinazar2021simdram}.
    By storing constant values (e.g., all-zeros or
    all-ones) as a control row, the majority operation can also implement Boolean logic such as AND and OR~\cite{seshadri2017ambit, fcdram, manyrowact}. Illustrated in Fig.~\ref{fig:dcc}, inversion has been proposed using dedicated dual-contact cells~\cite{seshadri2017ambit}. As an alternative, leveraging the realization of senseamps as SRAM-style crossed inverters as shown in the figure, inversion has also been realized by crossing the row buffer to the neighboring subarray~\cite{fcdram,drambender}.
    These operations process an entire row (e.g., typically 512- to 8192-bits)
    in a single activation cycle, providing massive bit-parallelism.
     
    \paragraph{Horizontal shifting}
    Recent work has proposed the ability to shift data laterally
    within a DRAM row by exploiting charge sharing between adjacent
    bitlines~\cite{tegge2026shiftingindram} or using dedicated logic within the row buffer~\cite{li2017drisa}.
    This enables rotation and shift operations without reading data
    out of the array, a capability critical for cryptographic
    permutations that rely on bitwise rotations for diffusion.

    \subsection{Trusted Execution Environments and Memory Encryption}
    \label{sec:bg-tee}
     
    A trusted execution environment (TEE) provides hardware-enforced isolation for sensitive computation, ensuring the confidentiality and integrity of code and data even in the presence of a compromised operating system, hypervisor, or co-resident tenant~\cite{costan2016intel,kaplan2016amd}.
    Intel SGX~\cite{costan2016intel} isolates application-level \emph{enclaves}, whose protected pages reside in the Enclave Page Cache (EPC), while AMD SEV~\cite{kaplan2016amd} provides VM-level isolation by encrypting each virtual machine's memory with a unique key. 
    
    Both architectures assume that trusted execution occurs within the processor. However, data leaving the processor to traverse the memory bus and reside in DRAM is exposed to physical attacks, including bus probing, cold-boot extraction, and active tampering. To mitigate these threats, Intel Total Memory Encryption (TME)~\cite{intel-tme} and AMD Secure Memory Encryption (SME) encrypt all data transferred between the processor and DRAM using AES-XTS, with the encryption engine integrated into the memory controller.

    This CPU-side encryption architecture creates significant complications for PIM execution. Several secure near-memory processing architectures have been proposed to address this limitation by extending the trusted computing base beyond the processor. SecNDP introduces a secure near-data processing architecture with dedicated trusted logic on the memory side for secure execution~\cite{xiong2022secndp}. SSS-DIMM similarly incorporates trusted execution capabilities within a secure smart DIMM~\cite{SSS-DIMM-TPDS25}. More recently, MemClave demonstrates secure memory-resident execution using commodity UPMEM near-memory processors~\cite{choudhari2026memclave,UPMEM-Hotchips19,UPMEM-Access22}. Collectively, these architectures recognize that secure PIM execution requires trusted execution environments in both the processor and memory while treating the memory bus as untrusted.

The use of charge-sharing PIM computation to build trusted execution environments is much more limited. Existing in-memory encryption approaches, including secure near-memory TEE proposals, are predominantly AES-based, typically employing AES-XTS for memory encryption~\cite{xiong2022secndp,SSS-DIMM-TPDS25,choudhari2026memclave,AESPIM}. Although AES has been implemented using PIM operations~\cite{AESPIM}, it incurs substantial execution overhead. Furthermore, AES-XTS provides confidentiality but does not provide integrity, freshness, or replay protection. Secure communication of the shared AES key over an untrusted memory bus is also challenging, and prior proposals typically rely on dedicated hardware or authenticated communication mechanisms to address this requirement~\cite{xiong2022secndp,SSS-DIMM-TPDS25,choudhari2026memclave,AESPIM}. Prior work also notes that classical public-key techniques, such as elliptic-curve cryptography (ECC), can be used to establish the shared AES key, but are prohibitively expensive to implement using charge-sharing PIM operations~\cite{AESPIM}.
     
\paperName{} addresses both limitations through cryptography-PIM co-design. Specifically, we co-design the LightSaber post-quantum KEM and ASCON-AEAD for efficient execution using PIM operations. Together, these primitives establish and maintain shared cryptographic state across cooperating processor- and memory TEEs, providing authenticated communication over an untrusted memory bus while enabling trusted execution within a unified distributed TCB.

    \subsection{Lattice-Based Key Encapsulation: Saber}
    \label{sec:bg-saber}

NIST ultimately standardized the Module-Lattice-Based Key-Encapsulation Mechanism (ML-KEM, formerly \textit{Kyber})~\cite{fips203}. The \textit{Saber} family~\cite{saber} was also a finalist in the NIST post-quantum cryptography standardization process. 

Although both ML-KEM and ECC rely on arithmetic modulo non-power-of-two primes, which requires modular reduction operations that are difficult to implement efficiently in charge-sharing PIM~\cite{AESPIM}, LightSaber achieves the same targeted post-quantum security levels as ML-KEM using modular reductions by powers of two ($q=8192=2^{13}$, $p=1024=2^{10}$). These reductions can be implemented as discarding high-order rows, while rounding operations reduce to simple row index reordering. When horizontal data movement is available~\cite{li2017drisa,tegge2026shiftingindram}, these operations map naturally to PIM execution. As a result, \paperName{} selects \emph{LightSaber}, the smallest variant of the Saber family.

LightSaber operates over the polynomial ring $R_q = \mathbb{Z}_q[X]/(X^{256} + 1)$ with a module rank of~2 (a $2 \times 2$ public matrix~$\mathbf{A}$ and $2 \times 1$ secret vector~$\mathbf{s}$ whose coefficients lie in $[-5, +5]$). The KEM consists of three operations: \emph{KeyGen} generates a public/private key pair; \emph{Encaps} produces a shared secret and a ciphertext; and \emph{Decaps} recovers the shared secret from the ciphertext using the private key.

    \subsection{ASCON Lightweight Authenticated Encryption}
    \label{sec:bg-ascon}
     
    ASCON is an authenticated encryption with associated data  scheme selected by NIST as the standard for lightweight cryptography~\cite{Ascon}. ASCON operates on a 320-bit internal state organized as five 64-bit words $(x_0, x_1, x_2, x_3, x_4)$. The state is divided into a \emph{rate} portion (the first $r$~bits, through which plaintext is absorbed and ciphertext is extracted) and a \emph{capacity} portion (the remaining $c$~bits, which are never directly exposed). ASCON uses $r = 64$ and $c = 256$, providing 128-bit security via $\min(k,\, \lfloor c/2 \rfloor,\, tag) = \min(128, 128, 128)$.
     
    The core of ASCON is its permutation, which consists of three layers per round: (i)~\emph{constant addition}: XOR a round constant into one word (1~operation); (ii)~\emph{substitution}: a 5-bit S-box applied bitwise across all 64 column positions, implemented as a sequence of XOR, AND, and NOT on the five words; and (iii)~\emph{linear diffusion}: each word is XORed with two rotated copies of itself .
    
    ASCON provides two substantial improvements over AES for \paperName{}. First, ASCON is an authenticated encryption scheme by design, whereas conventional AES-XTS memory encryption provides confidentiality alone and requires separate mechanisms to ensure integrity, authenticity, and replay protection.  Second, the SubBytes transformation step of AES, realized through either lookup tables or complex finite-field logic, is difficult to implement efficiently using bulk-bitwise DRAM operations and dominates execution time in charge-sharing PIM implementations~\cite{AESPIM}. In contrast, every operation in the ASCON permutation consists solely of XOR, AND, NOT, and fixed rotations, eliminating table lookups, finite-field arithmetic, and data-dependent branches.
     
    ASCON-128a is a throughput-optimized variant that doubles the rate
    to $r = 128$ (absorbing two words per block) at the cost of two
    additional rounds per block (8 vs.\ 6), yielding ${\sim}1.5\times$
    higher throughput than ASCON-128 on long messages while maintaining the same
    128-bit security target under a per-nonce data volume bound.

    \subsection{DRAM-Intrinsic Entropy Sources}
    \label{sec:bg-puf}
     
    A hardware root of trust requires a source of physical randomness
    that cannot be predicted, cloned, or extracted by software.
    DRAM cells provide such a source through two complementary
    mechanisms.
     
    \paragraph{Physical Unclonable Functions (PUF)} 
    The power-up initialization pattern of DRAM cells is influenced by nanoscale manufacturing variations (threshold voltage mismatch, capacitor geometry) that differ between chips.

    \paragraph{True Random Number Generation (TRNG)}
    An alternative approach exploits controlled row conflicts~\cite{uac-trng}: by simultaneously activating multiple rows in the same subarray, the charge on shared bitlines enters a metastable state whose resolution is governed by thermal noise. Because the entropy source is the physical charge dynamics of the DRAM cells themselves, the derived secrets are unique to each die and cannot be replicated on another module.


\section{The \paperName{} Trusted Computing Architecture}
\label{sec:YAVIN}




\begin{figure}
    \centering
    \begin{subfigure}[b]{.55\linewidth}
        \includegraphics[height=1.5in]{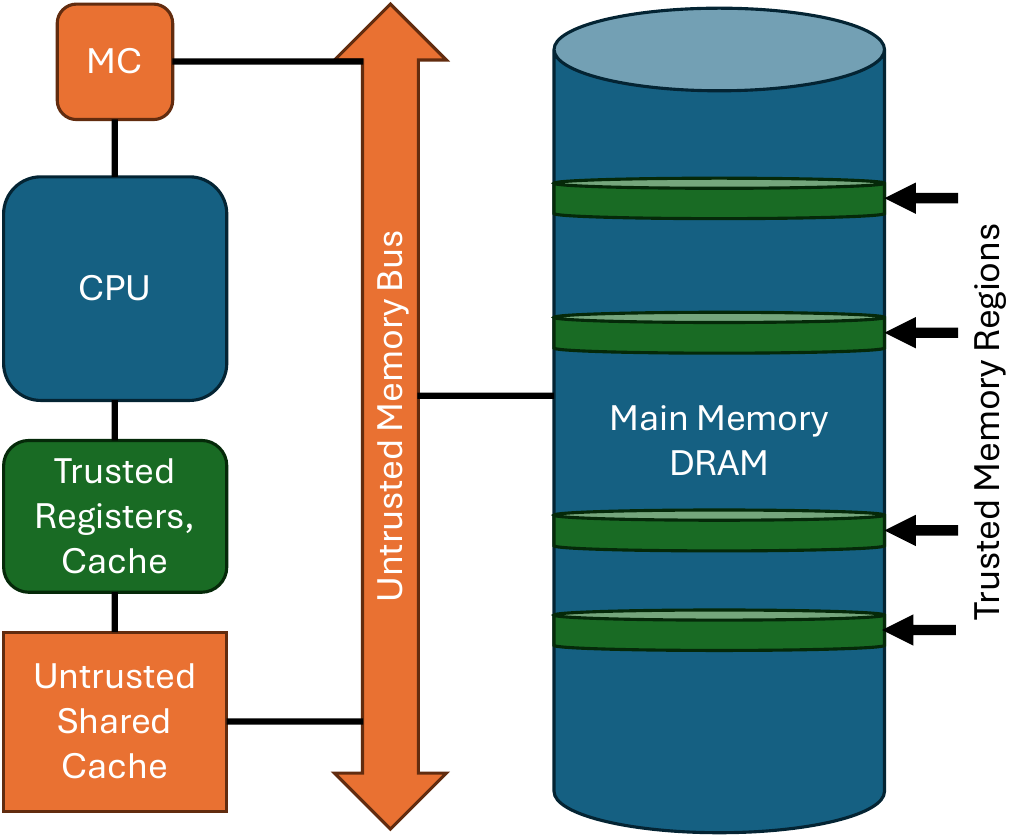}
        \caption{Architecture assumptions.}
        \label{fig:arch-assumptions}
    \end{subfigure}
    \begin{subfigure}[b]{.43\linewidth}
        \includegraphics[height=1.5in]{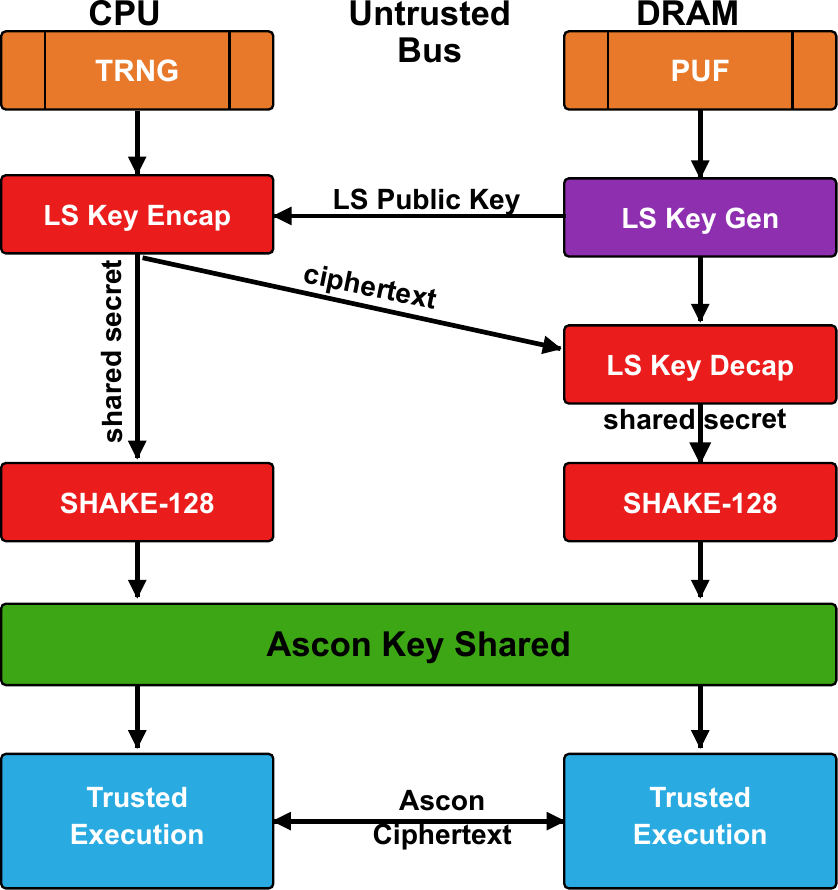}
        \caption{Trust establishment flow.}
        \label{fig:trust-establishment}
    \end{subfigure}
    \caption{\paperName{} TCB overview.}
    \label{fig:tcb-overview}
    \vspace{-.1in}
\end{figure}

The \paperName{} Architecture leverages the foundations established in Section~\ref{sec:background} to realize trusted execution in both the processor and memory, enabling PIM acceleration in the context of trusted multi-tenant execution.
\paperName{} adopts the security model of prior trusted execution environments~\cite{costan2016intel,kaplan2016amd,intel-tme}, including proposals for secure near-memory processing~\cite{xiong2022secndp,SSS-DIMM-TPDS25}. As illustrated in Fig.~\ref{fig:tcb-overview}, \paperName{} extends this model to charge-sharing PIM without requiring dedicated near-memory processing hardware embedded in DRAM. Instead, as shown in Fig.~\ref{fig:arch-assumptions}, trusted execution is permitted both within the CPU and within the memory device (blue), with isolated trusted storage 
(green) connected through an untrusted memory controller (MC) and bus (orange).  Consequently, \paperName{} ensures that only ciphertext traverses the memory bus. The MC is responsible for issuing PIM commands to the memory device. Although these commands reveal the requested cryptographic algorithm, they never expose plaintext or cryptographic keys.

Interoperable trusted execution is established through a shared symmetric key that lets both the CPU and the memory decrypt, process, and re-encrypt protected data. \paperName{} protects this data channel with the Ascon-AEAD standard~\cite{Ascon}, whose 128-bit key the two parties agree upon by running the LightSaber KEM~\cite{lightsaber}. Fig.~\ref{fig:trust-establishment} illustrates the flow.
 
The memory LightSaber key-pair generation process seeds the private key using a DRAM PUF~\cite{DRAM-PUF} and creates the public key which can be accessed over the untrusted bus. The CPU leverages its own randomness from the TRNG. The CPU then \emph{encapsulates} against that public key, an operation that simultaneously produces a random shared secret and a ciphertext, storing the ciphertext in memory. The memory \emph{decapsulates} the ciphertext to recover the identical shared secret. While the public key and the ciphertext cross the memory bus, neither reveals the shared secret without breaking the underlying Module-LWR security assumption. 
Both CPU and DRAM-PIM then derive the session key locally by conditioning the 256-bit shared secret through SHAKE-128 and retain the leading 128 bits as the Ascon key. Hashing ensures the key is a uniform function of the entire shared secret. 

\begin{figure}[tbp]
    \centering
    \includegraphics[width=\linewidth]{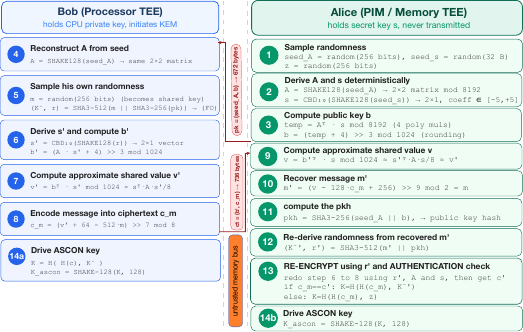}
    \caption{LightSaber flow in DRAM}
    \label{fig:ascon-flow}
    \vspace{-.1in}
\end{figure}

\begin{figure}[tbp]
    \centering
    \includegraphics[width=\linewidth]{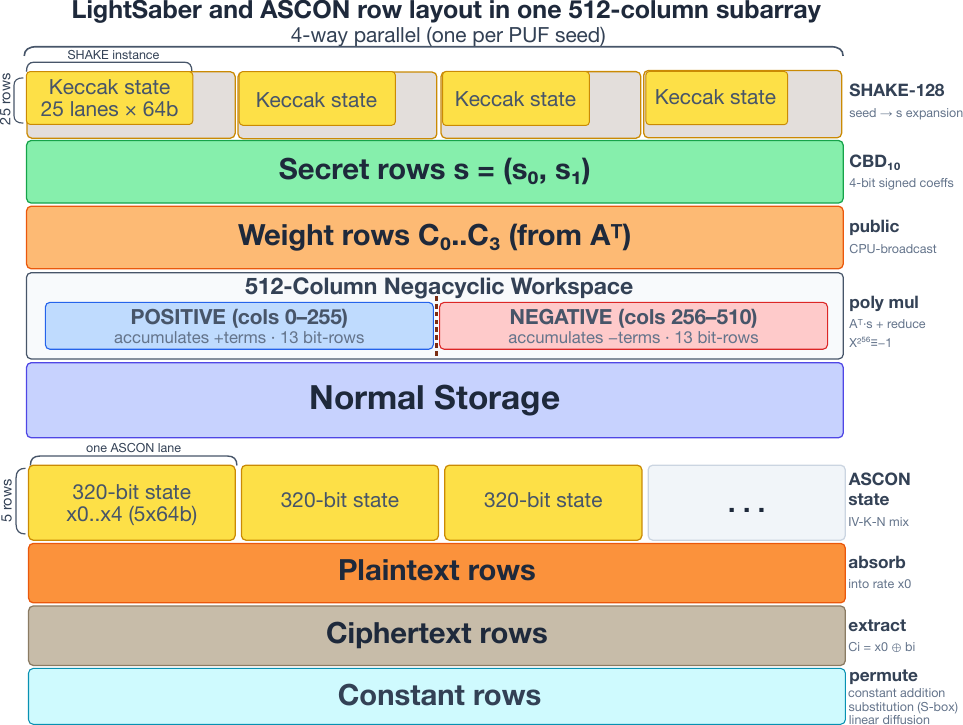}
    \caption{Row layout of the \paperName{} cryptographic engine within a single 512-column DRAM subarray. The upper region implements the one-time LightSaber key exchange. The lower region implements per-access ASCON-128 encryption.}
    \label{fig:yavin-layout}
    \vspace{-.1in}
\end{figure}

    \subsection{LightSaber design in PIM}


While LightSaber is well suited to charge-sharing PIM implementation, efficiently realizing the protocol still requires implementing key generation, encapsulation, and decapsulation directly within DRAM using bulk-bitwise operations. This section details the PIM implementation of \emph{key generation}. Decapsulation reuses the identical hardware building blocks (SHAKE-128 expansion, CBD sampling, and negacyclic polynomial multiply), differing only in which polynomials serve as inputs and in the choice of modular domain, i.e., $\bmod\; q = 8192$ for the 13-bit product and $\bmod\; p = 1024$ for the 10-bit rounded results. We therefore describe each of the building blocks once, in the KeyGen setting, noting that the same building blocks are reused throughout the remainder of the protocol. 

Key generation begins with a hardware root of trust derived from DRAM-intrinsic entropy (Section~\ref{sec:puf}), expands that entropy into the secret vector~$\mathbf{s}$ through parallel SHAKE-128 (Section~\ref{sec:shake}) and Centered Binomial Distribution (CBD) sampling (Section~\ref{sec:cbd}), and computes the public key~$\mathbf{b}$ through the polynomial product $\mathbf{A}^T \cdot \mathbf{s}$ (Section~\ref{sec:polymul}).
    
    \subsubsection{Hardware Root of Trust via DRAM Entropy}
        \label{sec:puf}
        
        A fundamental requirement of independent key generation in DRAM 
        is a root of trust that originates from the DRAM die itself, without relying on any secret provisioned by software or transmitted from the processor. \paperName{} derives this root of trust from the physical randomness inherent in DRAM cells~\cite{DRAM-PUF}, using a mechanism inspired by the QUAC-TRNG scheme~\cite{puf}.

\begin{figure}[tbp]
    \centering
    \includegraphics[width=\linewidth]{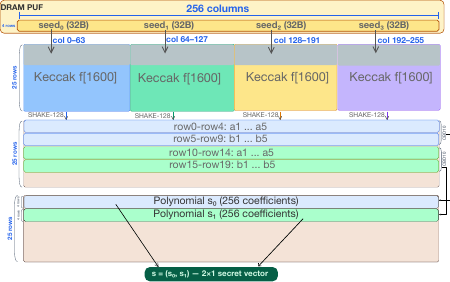}
    \caption{4-way parallel SHAKE-128 and $\textsc{CBD}_{10}$, generating vector \textbf{s}}
    \label{fig:shake128-cbd-layout}
    \vspace{-.2in}
\end{figure}

    \subsubsection{Seed Expansion via Parallel SHAKE-128}
    \label{sec:shake}
        Generating the secret vector~$\mathbf{s}$ requires a large volume of pseudorandom bits: the $\text{CBD}_{10}$ sampler consumes 10~bits per coefficient, and a $2 \times 1$ vector of 256-coefficient polynomials demands $2 \times 256 \times 10 = 5{,}120$ random bits in total. SHAKE-128, the extendable-output function built on the Keccak-$f$[1600] permutation, is the standard mechanism for expanding a short seed into an arbitrarily long pseudorandom stream. Its internal state is 1600~bits, organized as a $5 \times 5$ grid of 64-bit \emph{lanes} (25~lanes total), and its output rate is 1344~bits (168~bytes) per squeeze. A single LightSaber instance would therefore require $\lceil 5120 / 1344 \rceil = 4$ sequential SHAKE-128 squeeze operations to produce enough output, with each squeeze invoking a full 24-round Keccak permutation.
         
        \paragraph{Four-way parallelism from PUF entropy}
        The DRAM PUF provides sufficient entropy to generate four independent 32-byte seeds, allowing four SHAKE-128 instances to execute concurrently rather than sequentially. As shown in Fig.~\ref{fig:shake128-cbd-layout}, each instance occupies an independent group of 25 stacked Keccak lanes. Since every instance performs the same permutation, the same sequence of bulk-bitwise row activations is applied to all four groups simultaneously. Each instance produces 1344 bits in a single squeeze, yielding 5,376 bits in total, exceeding the 5,120 bits required for one LightSaber key generation. Only 256 bits of each 512-bit row are used intentionally during this stage. This is both sufficient to generate the required 256-bit secret while reserving the remaining capacity for subsequent phases of the algorithm.

    \begin{algorithm}[t]
    \caption{Bulk-Bitwise CBD$_{10}$ Sampling using PIM}
    \label{alg:pim-cbd}
    \begin{algorithmic}[1]
    \Require 10 DRAM rows $(a_1,\ldots,a_5,\, b_1,\ldots,b_5)$, from SHAKE-128 output
    \Output 4 DRAM rows $(r_3,r_2,r_1,r_0)$ encoding 256 coefficients $X \in [-5,+5]$ in 4-bit two's complement
    
    \Function{Popcount5}{$x_1, x_2, x_3, x_4, x_5$} \Comment{$5 \to 3$-bit}
      \State $c_1 \!\gets\! \textsc{Maj}(x_1,x_2,x_3)$;\; $t_1 \!\gets\! \textsc{Xor3}(x_1,x_2,x_3)$
      \State $c_2 \!\gets\! \textsc{And}(x_4,x_5)$;\; $t_2 \!\gets\! \textsc{Xor}(x_4,x_5)$
      \State $s_0 \!\gets\! \textsc{Xor}(t_1,t_2)$;\; $c_3 \!\gets\! \textsc{And}(t_1,t_2)$
      \State $s_1 \!\gets\! \textsc{Xor3}(c_1,c_2,c_3) $
      \State $s_2 \!\gets\! \textsc{Maj}(c_1,c_2,c_3) $
      \State \Return $(s_2,\, s_1,\, s_0)$ \Comment{value = $4s_2 + 2s_1 + s_0 \in [0,5]$}
    \EndFunction
    
    \State $(sa_2, sa_1, sa_0) \gets \Call{Popcount5}{a_1, a_2, a_3, a_4, a_5}$
    \State $(sb_2, sb_1, sb_0) \gets \Call{Popcount5}{b_1, b_2, b_3, b_4, b_5}$
    
    \For{$i = 0$ \textbf{to} $2$}  \Comment{$S_a - S_b$}
      \State $r_i \gets \textsc{Xor3}\!\left(sa_i,\;\overline{sb_i},\; c_{i-1}\right)$
      \State $c_i \gets \textsc{Maj}\!\left(sa_i,\;\overline{sb_i},\; c_{i-1}\right)$
    \EndFor
    \State $r_3 \gets \overline{c_2}$ \Comment{sign bit}
    
    \State \Return $(r_3,r_2,r_1,r_0)$ \Comment{$X = -8r_3 + 4r_2 + 2r_1 + r_0$}
    \end{algorithmic}
    \end{algorithm}

\begin{figure}[tbp]
    \centering
    \includegraphics[width=\linewidth]{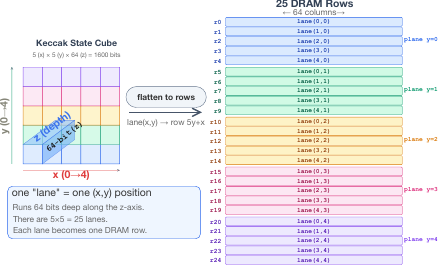}
    \caption{Keccak State: 5×5×64 Cube → 25 DRAM Rows}
    \label{fig:keccak_cube_to_rows}
    \vspace{-.2in}
\end{figure}
        \paragraph{Keccak permutation in PIM}

        Each 1600-bit Keccak state instance is stored as a stack of 25~DRAM rows in their own 64-bit column groups where each plane is stacked as five lanes as illustrated in Fig.~\ref{fig:keccak_cube_to_rows}. Thus, the lane at grid position $(x, y)$ is mapped to row index $5y + x$. 
        Operations that Keccak defines per-lane become operations on individual DRAM rows, while operations along the 64-bit lane axis become horizontal data movement within a row. Throughout the Keccak permutation, bit rotations in the lane are implemented using masking and horizontal shift operations~\cite{tegge2026shiftingindram,li2017drisa}. There are five steps in each permutation.
         
        The first step, $\theta$, provides inter-column diffusion. For each column~$x$, it XORs the lanes of that column into a 64-bit parity $C[x] \leftarrow \bigoplus_{y=0}^{4} \mathcal{S}[r_{x+5y}]$, forms a diffusion term $D[x] \leftarrow C[x{-}1] \oplus \mathrm{ROT}(C[x{+}1], 1)$, where $\mathrm{ROT}$ is a horizontal rotation. $D[x]$ is then XORed back into all five lanes of column~$x$. 

        The second step, $\rho$, rotates each of the 25 lanes, i.e. rows, by a fixed, row-specific offset ranging from 0 to 63~bits. 
         
        The third step, $\pi$, permutes the lanes according to the mapping $(x, y) \mapsto (y,\; 2x + 3y \bmod 5)$. This is accomplished by logical row reordering 
        incurring no data movement.
         
        In the fourth step, $\chi$, within each of the five planes (the five rows sharing a $y$ coordinate), it updates each row as $\mathcal{S}[x][y] \gets \mathcal{S}[x][y] \oplus (\overline{\mathcal{S}[x{+}1][y]} \wedge \mathcal{S}[x{+}2][y])$, where the two operand rows are the next two lanes in the same plane.
         
        The fifth step, $\iota$, XORs a round-dependent constant into a single lane (row), breaking the rotational symmetry that would otherwise make every round identical.

         
         

         
         
        After 24 rounds Keccak permutations, the four output blocks are partitioned to supply the CBD sampling, as in Fig \ref{fig:shake128-cbd-layout}.

    \subsubsection{Secret Sampling via Centered Binomial Distribution}
    \label{sec:cbd}
    
    
    Algorithm~\ref{alg:pim-cbd} describes the full $\text{CBD}_{10}$ sampler (10 input bits per coefficient) using three stages of column-parallel computation, producing 256 outputs simultaneously. The four parallel SHAKE-128 outputs are partitioned into 10-bit column-wise groups labeled $a_1$ to $a_5$ and $b_1$ to $b_5$. There are 256 column groups across those ten rows with some unneeded rows of parallel SHAKE-128 outputs discarded. The goal is to take the difference of the ones count for each column of $a$ and $b$ to determine an integer from -5 to 5 represented column-wise in four DRAM rows $(r_3, r_2, r_1, r_0)$ encoding each coefficient in 4-bit two's complement as the secret polynomial of $\mathbf{s}$.
    
    The \textsc{Popcount5} function (lines~1--7) sums five input bits to a 3-bit count $(s_2, s_1, s_0)$ satisfying $4s_2 + 2s_1 + s_0 = \sum_{i=1}^{5} x_i \in [0, 5]$. We modify the majority logic summation from SIMDRAM~\cite{hajinazar2021simdram} and tune the operation using parallel reduction to minimize the PIM operation count for Popcount5. Lines~10--14 compute $X = S_a - S_b$ by evaluating $S_a + \overline{S_b} + 1$ to produce $r_3$..$r_0$. These are combined with the coefficients of $A$ in the next step. As $A$ is a $2\times1$ vector we execute CBD$_10$ twice to produce a similar vector for $\mathbf{s}$ (see Fig.~\ref{fig:shake128-cbd-layout}).


\begin{figure}[tbp]
    \centering
    \includegraphics[width=\linewidth]{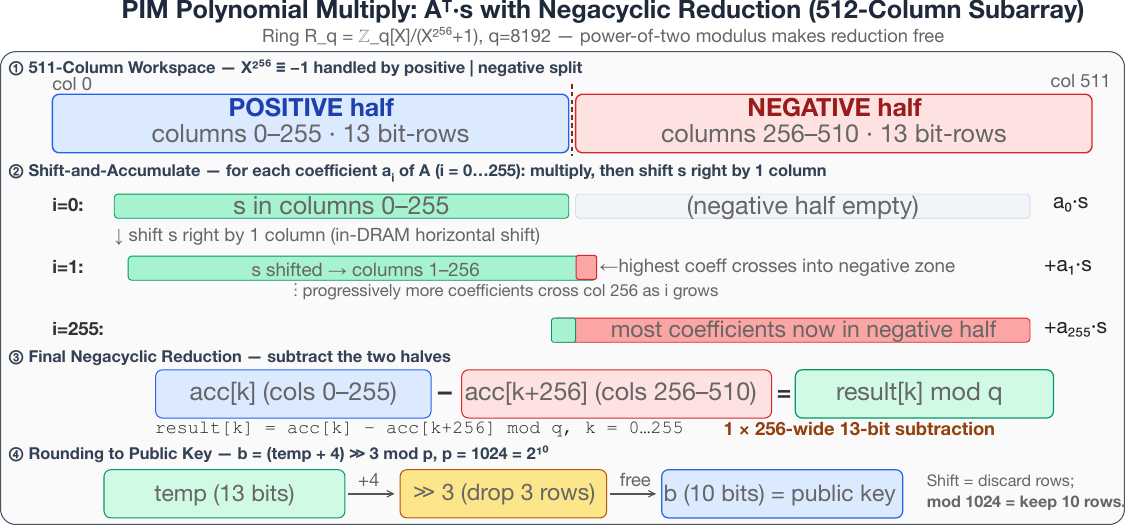}
    \caption{PIM polynomial multiplication for $\mathbf{A}^T\!\cdot\!\mathbf{s}$ in a 512-column subarray: a shift-and-accumulate loop places positive product terms in columns 0--255 and, exploiting $X^{256}\equiv-1$, lets high-degree terms cross into the negative half (columns 256--510) as $\mathbf{s}$ is shifted right; a single 256-wide subtraction then completes the negacyclic reduction}
    \label{fig:polymul_subarray}
    \vspace{-.1in}
\end{figure}

    \subsubsection{Polynomial multiply}
    \label{sec:polymul}
        The dominant computational cost in LightSaber KEM is the matrix--vector product $\mathbf{b} = \mathbf{A}^T \cdot \mathbf{s} + 4 \gg 3 \bmod p$, which requires four polynomial multiplications in the ring $R_q = \mathbb{Z}_q[X]/(X^n+1)$ with $q = 8192$ and $n = 256$. Each multiplication combines a public polynomial from~$\mathbf{A}$ (256 coefficients, 13~bits each) with the secret polynomial from~$\mathbf{s}$ (256 coefficients in $[-5,+5]$, computed in the last step\footnote{Because $\mathbf{A}$ is public it could originate in the processor, which uses its coefficients to precompute the multiply-accumulate weights and then store them in DRAM to support the PIM execution.}. 
        We decompose each multiplication into bulk-bitwise operations exploiting two structural properties: (i)~the secret coefficients are small, limiting the number of partial products to four (one per bit of magnitude plus one sign-correction term), and (ii)~all 256 coefficient positions are independent and stored as columns within a DRAM row, enabling all positions to be processed simultaneously with PIM commands.

        \begin{algorithm}[tbp]
        \caption{PIM Polynomial Multiplication for $\mathbf{A}^T\!\cdot\!\mathbf{s}$}
        \label{alg:pim-polymul}
        \begin{algorithmic}[1]
        \Require Matrix $\mathbf{A} \in R_q^{2\times 2}$, secret vector $\mathbf{s} \in R_q^{2\times 1}$ with coefficients in $[-5,+5]$, where $R_q = \mathbb{Z}_q[X]/(X^n+1)$, $q\!=\!8192$, $n\!=\!256$
        \Output Public key component $\mathbf{b} \in R_p^{2\times 1}$, $p = 1024$
        
        \Statex
        \Statex {\textbf{Phase 1: CPU precomputes coefficient weights}} 
        \For{each polynomial entry $A_{ij}$ of $\mathbf{A}^T$}
          \For{each coefficient $x \in A_{ij}$} \Comment{$x \in [0, q)$}
            \State $C_0 \gets x$ \Comment{weight of $D[0]$}
            \State $C_1 \gets 2x \bmod q$ \Comment{weight of $D[1]$}
            \State $C_2 \gets 4x \bmod q$ \Comment{weight of $D[2]$}
            \State $C_3 \gets (q - 8x) \bmod q$ \Comment{sign-corrected weight}
          \EndFor
          \State write $C_0, C_1, C_2, C_3$ to DRAM \Comment{13 rows $\times$ 256 cols}
        \EndFor
        
        \Statex
        \Statex \textbf{Phase 2: PIM shift, accumulate, negacyclic reduction}
        \Statex \Comment{Workspace: 512 cols $=$ 256 pos. $\|$ 256 neg.}
        \Statex \Comment{Reduct. rule: $X^{256} \equiv -1$ in $R_q = \mathbb{Z}_q[X]/(X^{256}+1)$}
         
        \State Init. 512-col $\textit{acc}[0\!:\!511] \gets 0$ \Comment{13 rows $\times$ 512 cols}
        \State Place $s_j$ coefficients in columns $0\!:\!255$ of workspace
         
        \For{$i = 0$ \textbf{to} $255$} \Comment{for each coefficient of $A_{ij}$}
          \State $D[3\!:\!0]$ is 4-bit signed repr.\ of $s_j$'s cur. col. alignment        \For{bit $b = 0$ \textbf{to} $2$} \Comment{magnitude bits}
            \State $\textit{partial} \gets \textsc{And}(C_i^{(b)},\; D[b])$ \Comment{select weight} 
            \State $\textit{acc} \gets \textit{acc} + \textit{partial}$ \Comment{13-bit add via MAJ chain}
          \EndFor
          \State $\textit{partial}_3 \gets \textsc{And}(C_i^{(3)},\; D[3])$ \Comment{sign-weighted term}
          \State $\textit{acc} \gets \textit{acc} + \textit{partial}_3$
          \State \textbf{Shift} $s_j$ right by 1 column in DRAM 
          \Statex \Comment{After shift: $s_j$ occupies columns $i{+}1$ to $i{+}256$}
          \Statex \Comment{Products at column $\ge 256$ land in negative zone}
        \EndFor
         
        \Statex
        \Statex \textbf{Phase 3: Negacyclic subtraction and rounding}
        \State $\textit{pos}[0\!:\!255] \gets \textit{acc}[0\!:\!255]$ \Comment{left half: positive terms}
        \State $\textit{neg}[0\!:\!254] \gets \textit{acc}[256\!:\!510]$ \Comment{right half: negative terms}
        \State $\textit{temp} \gets \textit{pos} - \textit{neg} \bmod q$ \Comment{negacyclic reduction done}
        \State $\mathbf{b} \gets (\textit{temp} + 4) \gg 3 \bmod p$ \Comment{round: truncate low 3 bits}
        
        \Statex
        \State \Return $\mathbf{b}$ \Comment{$2\!\times\!1$ vector, 10 bits $\times$ 256 coefficients}
        \end{algorithmic}
        \end{algorithm}



        \paragraph{CPU--PIM partitioning}
        Algorithm~\ref{alg:pim-polymul} splits the computation across the processor and DRAM-PIM. In Phase~1 (lines~1--9), the processor precomputes four \emph{coefficient weight rows} for each polynomial entry of $\mathbf{A}^T$: $C_0 = x$, $C_1 = 2x \bmod q$, $C_2 = 4x \bmod q$, and $C_3 = (q - 8x) \bmod q$, where $x$ is a coefficient of~$\mathbf{A}$. These weights correspond to the positional values of the 4-bit signed representation of the secret coefficient~$n \in [-5,+5]$: a coefficient $n$ with binary digits $D[3{:}0]$ satisfies $x \cdot n \equiv \sum_{i=0}^{3} C_i \cdot D[i] \pmod{q}$. Precomputing these weights on the CPU avoids expensive multiplication execution time in PIM\footnotemark[1]; the CPU then stores all four weight rows (each 13~bits $\times$ 256~columns) to DRAM.
         
        \paragraph{Bit-serial multiply-accumulate}
        In Phase~2 (lines~10--21), for each magnitude bit $D[i]$ ($i = 0, 1, 2$) of the secret coefficient, a row-wide \textsc{And} selects the corresponding weight row~$C_i$ at every column position where $D[i] = 1$ (line~15); the selected partial product is then accumulated into a 13-bit running sum via bit-serial addition built from \textsc{Maj} gates~\cite{hajinazar2021simdram} (line~16). 
        The sign bit $D[3]$ is handled separately: its weight~$C_3$ is constructed so that adding it to a negative accumulator and subtracting at the end correctly implements two's-complement arithmetic modulo~$q$.


        \paragraph{Negacyclic reduction via shift-and-accumulate} 
        Polynomial multiplication in $R_q = \mathbb{Z}_q[X]/(X^n + 1)$ with $n = 256$ differs from ordinary convolution in one critical respect: any product term whose degree reaches or exceeds~$n$ wraps around with a sign change, because $X^n \equiv -1$ in the quotient ring. Concretely, when multiplying coefficient~$a_i$ of~$\mathbf{A}$ by coefficient~$s_j$ of~$\mathbf{s}$, the contribution lands at degree~$i + j$. If $i + j < 256$, the contribution is added to the result; if $i + j \ge 256$, it must be \emph{subtracted} from degree $i + j - 256$. 
        We handle this reduction entirely within the DRAM subarray by allocating a \emph{512-column workspace}---twice the polynomial degree---partitioned into a left half (columns $0$--$255$, accumulating positive terms) and a right half (columns $256$--$510$, accumulating negative terms). 
        
        The secret polynomial~$\mathbf{s}$ is initially placed in columns $0$--$255$. For each successive coefficient~$a_i$ of~$\mathbf{A}$, the PIM performs the bit-serial multiply-accumulate described above across all 512 columns, then \emph{shifts~$\mathbf{s}$ right by one column} using an in-DRAM horizontal shift operation~\cite{tegge2026shiftingindram,li2017drisa}. After the shift, $\mathbf{s}$ occupies columns~$i{+}1$ through $i{+}256$: its lower-order coefficients remain in the left (positive) half, while the highest-order coefficient has crossed into column~$256$ and entered the right (negative) half. As~$i$ increases from~$0$ to~$255$, progressively more product terms land in the negative zone, exactly mirroring the $X^{256} \equiv -1$ reduction rule.
        After all 256 multiply-and-shift iterations, the workspace holds the complete unreduced product split into two aligned halves: columns $0$--$255$ contain the positive partial sum, and columns $256$--$510$
        contain the negative partial sum. The final reduction is a single 13-bit column-parallel subtraction:
        $$
        \textit{result}[k] = \textit{acc}[k] - \textit{acc}[k + 256] \bmod q, \quad k = 0, \ldots, 255
        $$

        \paragraph{Rounding}
        After the matrix--vector product yields $\textit{temp} = \mathbf{A}^T \cdot \mathbf{s} \bmod q$ in the 13-bit domain, line~25 of Algorithm~\ref{alg:pim-polymul} computes the rounded public key as $\mathbf{b} = (\textit{temp} + 4) \gg 3 \bmod p$. The right shift by 3 and reduction modulo $p = 1024 = 2^{10}$ are accomplished through logical reindexing, discarding the three rows representing the least-significant bits, thus truncating to the most significant ten rows. 

At this point the memory now holds a secret~$\mathbf{s}$ which should not be accessed by the CPU while allowing the CPU to access the corresponding published public key~$\mathbf{b}$ . The processor encapsulates a shared secret against this public key and stores a ciphertext which DRAM-PIM can decapsulate using the same polynomial-multiply and sampling primitives described above, adapted to the smaller $\bmod\; p$ domain. 

Decapsulation is hardened against chosen-ciphertext attacks through the Fujisaki--Okamoto (FO) transform, which requires two additional hash functions: SHA3-256 for deriving the shared key, i.e., the ASCON key, and the FO re-encryption seed and SHA3-512 for deriving the implicit-rejection randomness. After recovering the message~$m'$ computed in the processor, the PIM re-encrypts it and verifies that the result matches the received ciphertext. A mismatch triggers implicit rejection. Crucially, both SHA3-256 and SHA3-512 are built on the same Keccak-$f$[1600] permutation as SHAKE-128 (Section~\ref{sec:shake}). Our PIM implementation therefore reuses the identical 25-row state layout and five-step permutation already described
. Finally, at the end of this one-time exchange, after checking the authentication and deriving the 128-bit ASCON key, the CPU and memory both hold an identical 256-bit shared secret without it being revealed in plaintext over the untrusted memory bus.




     

    \subsection{PIM Ascon-AEAD design}
    
    With the shared ASCON key established by LightSaber, all subsequent data crossing the untrusted bus across the unified TEE is protected by ASCON-AEAD~\cite{Ascon}. Like the Keccak primitives above, ASCON's permutation uses only bitwise XOR, AND, NOT, and horizontal rotation. 
 
  \subsubsection{State layout}

    The Ascon-128 state is 320~bits, organized as five 64-bit words ($x_0, x_1, x_2, x_3, x_4$) as in Table~\ref{tab:ascon-state}.
    We map each word to one DRAM row in a stack similar to the lane organization in LightSaber, yielding a five-row state block initialized as follows:
     
    \begin{table}[h]
    \caption{ASCON state in DRAM}
    \label{tab:ascon-state}
    \renewcommand{\arraystretch}{1.15}
    \centering
    \footnotesize
    \begin{tabular}{|c|c|l|}
    \hline
    \textbf{Row} & \textbf{Word} & \textbf{Initial contents (64 bits)} \\
    \hline
    0 & $x_0$ & \texttt{0x80400c0600000000} (fixed IV) \\
    1 & $x_1$ & $K_0$ (key bits 127\,--\,64) \\
    2 & $x_2$ & $K_1$ (key bits 63\,--\,0) \\
    3 & $x_3$ & $N_0$ (nonce bits 127\,--\,64) \\
    4 & $x_4$ & $N_1$ (nonce bits 63\,--\,0) \\
    \hline
    \end{tabular}
    \end{table}
    \noindent
The Initialization Vector (IV) encodes the algorithm parameters: key size $k = 128$, rate $r = 64$, initialization rounds $a = 12$, and data-phase rounds $b = 6$. Words $x_1$ and $x_2$ store the 128-bit session key derived from LightSaber, while $x_3$ and $x_4$ store a 128-bit nonce. Because each ASCON state occupies only 64 bits of a 512-bit DRAM row, we instantiate eight parallel AEAD instances within each row, all sharing the same session key. A CPU-managed nonce counter is combined with a unique instance identifier to generate a distinct nonce for each AEAD instance. The nonce need not remain secret to preserve ASCON's security and therefore may be stored in plaintext and transmitted over the untrusted bus.

    \subsubsection{Permutation implementation}
    \label{sec:ascon-permutation}
     
    Each round of the ASCON permutation applies three layers sequentially. We detail the PIM implementation as follows:
     
    \paragraph{Layer~1: Constant addition.}
    A round-dependent constant is XORed into $x_2$:
    $x_2 \leftarrow x_2 \oplus \textit{RC}[\textit{round}]$.
    This requires a single bitwise XOR between the $x_2$ row and a pre-loaded constant row.

    \paragraph{Layer~2: Substitution.}
    A 5-bit S-box is applied bitwise across all 64 column positions simultaneously via three phases. Recall that unlike the AES S-box, which is based on inversion in a Galois field and is typically implemented using lookup tables, the ASCON S-box is constructed entirely from Boolean operations. This bitwise formulation maps naturally to DRAM-PIM, eliminating the need for lookup tables and enabling efficient bulk-bitwise execution.
    \textit{Phase~A1} (pre-mix) mixes three word pairs:
    $x_0 \leftarrow x_0 \mathrel{\oplus} x_4$,\;
    $x_4 \leftarrow x_4 \mathrel{\oplus} x_3$,\;
    $x_2 \leftarrow x_2 \mathrel{\oplus} x_1$.
    \textit{Phase~$\chi$} (nonlinear core) applies the only nonlinear operation, computing $t_i \leftarrow \overline{x_i} \wedge x_{(i+1)\bmod 5}$ for each word $i \in \{0,\ldots,4\}$, then updating $x_i \leftarrow x_i \mathrel{\oplus} t_{(i+1)\bmod 5}$. Here, $\bmod 5$ applies only to the word indices, implementing circular indexing from word 4 back to word 0.
    \textit{Phase~A2} (post-mix) completes the S-box:
    $x_1 \leftarrow x_1 \mathrel{\oplus}  x_0$,\;
    $x_0 \leftarrow x_0 \mathrel{\oplus} x_4$,\;
    $x_3 \leftarrow x_3 \mathrel{\oplus} x_2$,\;
    $x_2 \leftarrow \overline{x_2}$.

    \paragraph{Layer~3: Linear diffusion.}
    Each word is mixed with two rotated copies of itself,
    $x_i \leftarrow x_i \oplus (x_i \ggg r_{i,1}) \oplus (x_i \ggg r_{i,2})$,
    with rotation pairs
    $(19,28)$,
    $(61,39)$,
    $(1,6)$,
    $(10,17)$,
    $(7,41)$
    for $x_0$ through $x_4$, respectively.
    
    \subsubsection{AEAD operation flow}
    \label{sec:ascon-aead-flow}

    The full Ascon-128 AEAD procedure consists of four phases, each mapping to a specific sequence of PIM operations on the five-row state block. The initialization and finalization phases use the full 12-round permutation~$p^{12}$ for complete state mixing, while the data phase uses the cheaper 6-round permutation~$p^{6}$ between blocks for throughput.
 
    \paragraph{Phase~1: Initialization}
    Starting from the initial state $S_0 = (IV, K_0, K_1, N_0, N_1)$, apply the 12-round permutation $p^{12}$, then inject the key back into the state $S$ via XOR:
    $$
    S \leftarrow p^{12}(S_0), \quad x_3 \leftarrow  x_3 \mathrel{\oplus} K_0, \quad x_4 \leftarrow x_4 \mathrel{\oplus} K_1
    $$

    \paragraph{Phase~2: Associated data processing}

    Associated data (AD), i.e., authenticated but unencrypted data, is absorbed in 64-bit blocks, with a $p^{6}$ permutation applied after each block. After the final block, a single domain-separation bit is XORed into the state:
    $$
    x_0 \leftarrow x_0 \oplus \textit{padding}(\text{AD}), \quad S \leftarrow p^{6}(S), \quad x_4 \leftarrow x_4 \mathrel{\oplus} 1
    $$
    Padding appends a single \texttt{1} bit followed by zeros to complete the final 64-bit block. If no associated data is present, the absorption phase is skipped and only the domain-separation bit is applied before processing the plaintext.

    \paragraph{Phase~3: Plaintext encryption}
    \label{sec:plaintext-encryption}
    The plaintext is split into 64-bit blocks $b_1, b_2, \ldots, b_n$. For each block, the ciphertext is extracted by XORing the rate word~$x_0$ with the plaintext, the plaintext is then absorbed into the state, and a 6-round permutation is applied before the next block:
    $$
    C_i \leftarrow x_0 \oplus b_i, \quad x_0 \leftarrow x_0 \oplus b_i, \quad S \leftarrow p^{6}(S).
    $$ 
    For the last block, padding is applied and no permutation follows.
    
    ASCON-128a, the throughput-optimized variant, doubles the rate to 128~bits by absorbing through the first two state words $(x_0, x_1)$ reducing the capacity to 192~bits from the last three state words ($x_2, x_3, x_4$).  Plaintext is split into 128-bit blocks $B_1, B_2, \ldots, B_n$ such that 64-bit sub-blocks of $B_i = b_i^0, b_i^1$. To maintain 128-bit cryptographic strength with the lower capacity, a stronger 8-round permutation is applied between blocks doubling the rate with only a 33\% increase in permutations:
    $$
    C_i^{0} \leftarrow x_0 \oplus b_i^{0}, \qquad C_i^{1} \leftarrow x_1 \oplus b_i^{1},
    $$
    $$
    x_0 \leftarrow C_i^{0}, \quad x_1 \leftarrow C_i^{1}, \quad S \leftarrow p^{8}(S).
    $$

    \paragraph{Phase~4: Finalization}    
    The key is injected into the state through words $x_1, x_2$, a 12-round permutation is applied, the key is then reinjected into the state through words $x_3, x_4$, after which the 128-bit tag is extracted:
    \begin{align*}
    x_1 &\leftarrow x_1 \oplus K_0, \quad x_2 \leftarrow x_2 \oplus K_1 \nonumber \\
    S &\leftarrow p^{12}(S) \\
    x_3 &\leftarrow x_3 \oplus K_0, \quad x_4 \leftarrow x_4 \oplus K_1 \nonumber \\
    \textit{tag} & \leftarrow x_3 \| x_4 \quad (128~\text{bits}) \nonumber
    \end{align*}
The receiver recomputes the authentication tag and compares it with the received tag; any mismatch indicates that the ciphertext or associated data has been modified..

\subsubsection{Parallelism and nonce management}
\label{sec:ascon-nonce}
    Since ASCON state words are 64~bits wide, \paperName{} exploits memory row width, e.g., 512 bits, to pack \emph{eight independent ASCON instances}, achieving an $8\times$ increase in parallelism. The eight instances share a single session key~$K$, so \paperName{} ensures nonce uniqueness by partitioning the 128-bit nonce as $N = (\mathit{instance\_id} \,\|\, \mathit{ctr})$: a small instance-identifier field (distinguishing the eight instances) followed by a shared block counter maintained by the CPU. The instance identifier ensures that no two instances ever share a nonce even though they share the same key and global counter, while the monotonically increasing counter ensures that no instance repeats a nonce across successive blocks.

\subsection{Threat Model}
\paperName{} assumes the processor and the DRAM are trusted, including their trusted-execution logic, per-tenant isolated regions, and the secrets confined to them, whereas the memory bus and memory controller are untrusted. Adversaries such as co-resident tenants, a compromised OS/hypervisor, or attackers with physical access can observe all bus traffic and actively inject or replay messages on the bus.

Under this model \paperName{} ensures that only ciphertext traverses the bus: confidentiality of the one-time key exchange is guaranteed by the IND-CCA security of LightSaber. We assume a PUF-derived hardware root of trust and a manufacturing-time device identity that authenticates the initial public-key exchange. Consistent with standard TEE assumptions, physical side-channel attacks, cold-boot attacks, and denial-of-service attacks are outside the scope of this work.

\begin{table}[tbp]
\renewcommand{\arraystretch}{1.15}
\caption{Experimental platforms.}
\label{tab:exp-platform}
\centering
\footnotesize
\begin{tabular}{lcc}
\toprule
\textbf{Parameter} & \textbf{DDR3} & \textbf{DDR4} \\
\midrule
Memory standard            & DDR3-1333 & DDR4-2400 \\
Clock frequency            & 666\,MHz  & 1200\,MHz \\
Supply voltage             & 1.5\,V    & 1.2\,V \\
Channels, Ranks/channel                   & 2, 2         & 2, 2 \\
Devices/rank, Banks/chip               & 8, 8         & 8, 8 \\
Bus width                  & 64\,bits  & 64\,bits \\
Chip capacity              & 4\,Gb     & 8\,Gb \\
Row size                   & 1\,kB     & 1\,kB \\
Rows/bank                  & 65,536    & 65,536 \\
$t_{\mathrm{RCD}} = t_{\mathrm{RP}}$         & 13.5\,ns  & 14.2\,ns \\
$t_{\mathrm{RAS}}$         & 36\,ns    & 32.5\,ns \\
$t_{\mathrm{RC}}$          & 49.5\,ns  & 46.7\,ns \\
$t_{\mathrm{RFC}}$         & 160.5\,ns & 350\,ns \\
$t_{\mathrm{FAW}}$         & 30\,ns    & 21.7\,ns \\
\midrule
Processor                  & \multicolumn{2}{c}{Dual-core ARM Cortex-A72} \\
Clock frequency            & \multicolumn{2}{c}{1.2\,GHz} \\
\bottomrule
\end{tabular}
\vspace{-.1in}
\end{table}
\section{Experimental Evaluation}
\label{sec:results}
We evaluate YAVIN along 4 dimensions: (1)~the cost of Lightsaber key establishment, (2)~the throughput, latency, and energy efficiency of the ASCON cryptographic engine on the PIM, (3)~the scalability advantage over CPU-side TEEs in multi-tenant settings, (4)~end-to-end application performance under memory-intensive and compute-heavy workloads. 

Table~\ref{tab:exp-platform} summarizes the experimental platforms. We evaluate AES, LightSaber, and ASCON PIM implementations using NVMain-PIM configured with DDR3-1333 and DDR4-2400 timing parameters. Unless otherwise noted, all workloads execute within a single DRAM bank, and the reported latency and energy correspond to that bank. Processor-side experiments execute on a dual-core ARM Cortex-A72 operating at 1.2,GHz to measure LightSaber, ASCON, and the data reorganization and transposition overheads required to interface with the PIM execution model.

\subsection{\paperName{}'s space overhead}
Because each tenant establishes an independent secure channel, \paperName{} reserves a small, private region of one DRAM subarray per tenant to hold cryptographic state. We quantify this cost by component (LightSaber and ASCON); both primitives operate within the subarray's native 512-column width.

\paragraph{LightSaber region.}
Key generation and decapsulation perform polynomial multiplication in the ring $R_q = \mathbb{Z}_q[X]/(X^{256}+1)$, which uses the full 512-column workspace: 256 columns accumulate positive product terms and 256 accumulate negative terms, realizing the negacyclic reduction $X^{256}\equiv-1$ (Section~\ref{sec:polymul}). Storing the PUF-derived root of trust, the secret vector~$\mathbf{s}$, and the intermediate polynomial multiplication result requires approximately 120 rows. This region is used only during the one-time session handshake (${\approx}7{,}680$~bytes).

\paragraph{ASCON region.}
To make best use of the subarray, \paperName{} runs eight ASCON lanes in parallel: since one state word is 64~bits wide, eight independent instances take 512 columns, all sharing the session key but using distinct per-lane nonces (Section~\ref{sec:ascon-nonce}). The five-word state, round constants, and permutation scratch occupy up to 25 rows across those 512 columns (${\approx}1{,}600$~bytes), serving all eight lanes at once.

Summing the two regions, \paperName{} reserves \textbf{9{,}280~bytes (${\approx}9.1$\,KiB)}, per tenant per subarray. This is a conservative upper bound: because the LightSaber region is exercised only during the one-time key exchange, its 120 rows can be freed once the session is established. The per-tenant cost is thus a negligible fraction of DRAM device and scales linearly with the number of concurrent tenant sessions.



\subsection{Lightsaber KEM: Latency, and Energy}
\label{sec:eval-kem}

\begin{table}[t]
\centering
\renewcommand{\arraystretch}{1.1}
\caption{PIM side Lightsaber KEM Per-Step Latency and Energy over DDR3-1333 and DDR4-2400}
\label{tab:lightsaber-kem}
\footnotesize
\begin{tabular}{|l|l|c|c|}
\hline
& \textbf{Trace} & \textbf{DDR3} & \textbf{DDR4} \\
\hline
\multicolumn{4}{|c|}{\textit{Latency (ms)}} \\
\hline
\multirow{5}{*}{\rotatebox{0}{\textbf{KeyGen}}}
 & Seed expansion (SHAKE128)         &  6.09  &  6.04  \\ \cline{2-4}
 & Secret vector ($CBD_{10}$)            &  0.02 &  0.02 \\ \cline{2-4}
 & $A^T \cdot s$           & 394 & 391 \\ \cline{2-4}
 & Compute b     &  0.01 &  0.01 \\ \cline{2-4}
 & Compute PKH (SHA3-256) & 32.51  & 32.23  \\
\hline
\multirow{6}{*}{\rotatebox{0}{\textbf{Decaps}}}
 & calculate $v=b^{'^{T}}\cdot s$     &  85.96 &  85.24 \\ \cline{2-4}
 & Recover message (m') &  0.02 &  0.02 \\ \cline{2-4}
 & Re-derive $K'$, $r'$ (SHA3-512) &  6.13  &  6.08  \\ \cline{2-4}
 & Re-encrypt using r', get c' & 480 & 476 \\ \cline{2-4}
 & Authentication check & 97.53 & 96.7 \\ \cline{2-4}
 & Derive ASCON key & 6.09 & 6.04 \\
\hline
\multicolumn{4}{|c|}{\textit{Energy (mJ)}} \\
\hline
\multirow{5}{*}{\rotatebox{0}{\textbf{KeyGen}}}
 & Seed expansion (SHAKE128)         & 0.85   & 0.85   \\ \cline{2-4}
 & Secret vector ($CBD_{10}$)            & 0.003 & 0.003 \\ \cline{2-4}
 & $A^T \cdot s$           & 55.08  & 55.04  \\ \cline{2-4}
 & Compute b     & 0.001 & 0.001 \\ \cline{2-4}
 & Compute PKH (SHA3-256) & 4.48   & 4.48   \\
\hline
\multirow{6}{*}{\rotatebox{0}{\textbf{Decaps}}}
 & calculate $v=b^{'^{T}}\cdot s$     & 11.88  & 11.87  \\ \cline{2-4}
 & Recover message (m') & 0.002 & 0.002 \\ \cline{2-4}
 & Re-derive $K'$, $r'$ (SHA3-512) & 0.85   & 0.85   \\ \cline{2-4}
 & Re-encrypt using r', get c' & 66.96 & 66.91 \\ \cline{2-4}
 & Authentication check & 13.44 & 13.44 \\ \cline{2-4}
 & Derive ASCON key & 0.85   & 0.85   \\
\hline
\end{tabular}
\end{table}

\begin{table}[tbp]
\renewcommand{\arraystretch}{1.15}
\caption{ASCON Encryption Latency and Energy under different memory config (DDR3-1333 and DDR4-2400 ) compared with PIM AES.}
\label{tab:ascon-latency-energy}
\centering
\footnotesize
\begin{tabular}{|l|cc|cc|cc|}
\hline
\textbf{Data} &
\multicolumn{2}{c|}{\textbf{ASCON-128}} &
\multicolumn{2}{c|}{\textbf{ASCON-128a}} &
\multicolumn{2}{c|}{\textbf{AES-128}} \\
\cline{2-7}
\textbf{Size} & \textbf{ms} & \textbf{mJ}
& \textbf{ms} & \textbf{mJ}
& \textbf{ms} & \textbf{mJ} \\
\hline
\multicolumn{7}{|c|}{\textbf{DDR3-1333}} \\
\hline
64\,b  & 2.93 & 0.39 & 3.13 & 0.416 & 21.9   & 2.99 \\
128\,b & 3.53 & 0.47 & 3.13 & 0.416 & 21.9  & 2.99 \\
1\,Kb  & 11.7  & 1.56  & 8.60  & 1.14  & 174   & 24.0 \\
1\,Mb  & 9606  & 1277  & 6405  & 851   & 178995 & 24543 \\
\hline
\multicolumn{7}{|c|}{\textbf{DDR4-2400}} \\
\hline
64\,b  & 2.90 & 0.39 & 3.09 & 0.42 & 23.7   & 2.99 \\
128\,b & 3.48 & 0.47 & 3.09 & 0.42 & 23.7   & 2.99 \\
1\,Kb  & 11.6  & 1.56  & 8.51  & 1.14  & 189    & 23.9 \\
1\,Mb  & 9508  & 1277  & 6340  & 851   & 193904 & 24518 \\
\hline
\end{tabular}
\vspace{-.1in}
\end{table}

As LightSaber establishes the shared ASCON session key, it runs once per session and comprises two PIM-side phases---key generation (KeyGen) and decapsulation (Decaps)---plus a processor-side encapsulation whose cost is negligible. Table~\ref{tab:lightsaber-kem} reports per-step latency and energy across DDR3-1333 and DDR4-2400.
 
\paragraph{KeyGen}
Key generation completes in 429\,ms and 60.4\,mJ, of which the matrix--vector product $\mathbf{A}^T\!\cdot\mathbf{s}$ accounts for 91\% (391\,ms, 55.0\,mJ). This is expected, since it comprises four polynomial multiplications of 256 multiply-accumulate iterations each over 13-bit coefficients. The remaining steps are minor: SHAKE-128 seed expansion adds 6.04\,ms (1.5\%), while $\text{CBD}_{10}$ sampling (0.02\,ms) and the rounding to~$\mathbf{b}$ (0.01\,ms) are negligible. This step can be executed offline because it is a randomized, independent primitive that requires no prior interaction with the recipient, recipient state, or communication channel.

\begin{table*}[tbp]
\renewcommand{\arraystretch}{1.05}
\setlength{\tabcolsep}{1.5pt}
\caption{YAVIN Encryption Overhead on LLM Inference via DDR4-2400}
\label{tab:llm-overhead}
\centering
\footnotesize
\begin{tabular}{|l|r|r|r|r|r|r|r|}
\hline
\textbf{Metric} & \textbf{GPT-J 6B} & \textbf{GPT-OSS 20B} & \textbf{LLaMA-3.1 8B} & \textbf{LLaMA-3.3 70B} & \textbf{LLaMA-4 Scout} & \textbf{Mixtral 8x7B} & \textbf{Qwen3-VL 4B} \\
\hline
\multicolumn{8}{|c|}{\textbf{INT8}} \\
\hline
PIM GEMM (s)          & 124.9  & 36.5   & 171.7   & 1518.3  & 244.3   & 31.4    & 81.9 \\
CPU crypto (s) & 0.54 & 3.12 & 0.67 & 3.29 & 1.08 & 0.86 & 0.64 \\
Bs-Bp roundtrip (s) & 0.24 & 1.37 & 0.27 & 1.45 & 0.51 & 0.38 & 0.28 \\
Crypto time in PIM (s)  & 34 & 198.75 & 42.5 & 210  & 74.4 & 55 & 40.625\\ 
\textbf{YAVIN overhead}   & 27.85\% & 556.82\% & 25.30\% & 14.14\% & 31.11\% & 179.11\% & 50.73\% \\
\hline
\multicolumn{8}{|c|}{\textbf{INT32}} \\
\hline
PIM GEMM (s)   & 1777.6 & 527.3  & 2331.0  & 19983.2 & 3277.4  & 485.8   & 1311.3 \\
CPU crypto (s) & 2.15 & 12.47 & 2.65 & 13.18 & 4.66 & 3.43 & 2.57 \\
Bs-Bp roundtrip (s) & 0.94 & 5.48 & 1.16 & 5.79 & 2.05 & 1.51 & 1.13 \\
Crypto time in PIM (s)  & 136.88 & 795.00 & 168.75 & 840.00 & 296.88 & 218.75 & 163.75 \\ 
\textbf{YAVIN overhead}   & 7.87\% & 154.18\% & 7.40\% & 4.30\% & 9.26\% & 46.05\% & 12.77\% \\
\hline
\end{tabular}
\vspace{-.1in}
\end{table*}

\paragraph{Decapsulation} 
Decapsulation is the critical-path operation because it provides the CCA security. It completes in 670.2\,ms and consumes 93.9\,mJ. The dominant cost is the \emph{re-encryption} step required by the Fujisaki--Okamoto (FO) transform for CCA security: to verify the received ciphertext, the PIM must repeat the full encapsulation using the recovered message~$m'$, costing 390.9\,ms---identical to the $A^T \cdot s$ in the \textbf{KeyGen}, as expected, since it executes the same matrix--vector multiplication. The inner product $\mathbf{b'}^T \cdot \mathbf{s}$ used to recover the shared value takes 85.2\,ms, roughly $4.6\times$ faster than the full $\mathbf{A}^T \cdot \mathbf{s}$ product because it involves only two polynomial multiplications (a $1 \times 2$ vector--vector product) in the smaller 10-bit modular domain ($\bmod\; p = 1024$) rather than four multiplications in the 13-bit domain ($\bmod\; q = 8192$). The hash operations (SHA3-256 for the shared key, SHA3-512 for FO randomness re-derivation) together take 38.3\,ms, which is relatively modest to the polynomial arithmetic. The ciphertext authentication check adds 96.7\,ms, and the final derivation of the 128-bit ASCON key from the shared secret via SHAKE-128 adds 6.04\,ms.

\paragraph{Total one-time cost and amortization}
The complete key exchange (CPU Encaps + PIM Decaps) costs $T_{\mathrm{Enc\_CPU}} + 670.2$\,ms and $E_{\mathrm{Enc\_CPU}} + 93.9$\,mJ. This is a one-time session establishment cost. In terms of user experience, this is a nominal subsecond overhead to establish secure trusted execution. Once the shared ASCON key is in place, every subsequent memory access is protected by ASCON encryption.


\subsection{PIM ASCON v.s PIM AES}
\label{sec:eval-ascon}

We characterize the PIM-side ASCON engine in isolation. Table~\ref{tab:ascon-latency-energy} reports encryption latency and energy from a single 64-bit block up to 1\,Mb across different DDR configurations. Against AES-128, ASCON's bulk-bitwise operations are cheaper across all data sizes, and the higher-throughput ASCON-128a mode (128-bit rate, 8-round permutation) yields a further ${\sim}1.5\times$ improvement over ASCON-128.

Across both memory configurations, PIM ASCON consistently outperform AES-128 in PIM execution. On DDR4-2400, ASCON-128 achieves an \textbf{8.2$\boldsymbol{\times}$} latency speedup at 64\,b (2.90\,ms vs.\ 23.67\,ms), increasing to \textbf{20.4$\boldsymbol{\times}$} at 1\,Mb as initialization overhead amortizes. Energy follows the same trend, \textbf{7.7$\boldsymbol{\times}$} at 64\,b and \textbf{19.2$\boldsymbol{\times}$} at 1\,Mb. ASCON-128a further extends the advantage at large data sizes thanks to its doubled rate, reaching \textbf{28.8$\boldsymbol{\times}$} energy reduction over AES at 1\,Mb. These ratios remain consistent across DDR3-1333, confirming that the advantage is inherent to the algorithm's suitability for PIM rather than a memory-configuration artifact.

\subsection{FHE v.s \paperName{}}
\begin{figure}[tbp]
    \centering
    \includegraphics[width=\linewidth]{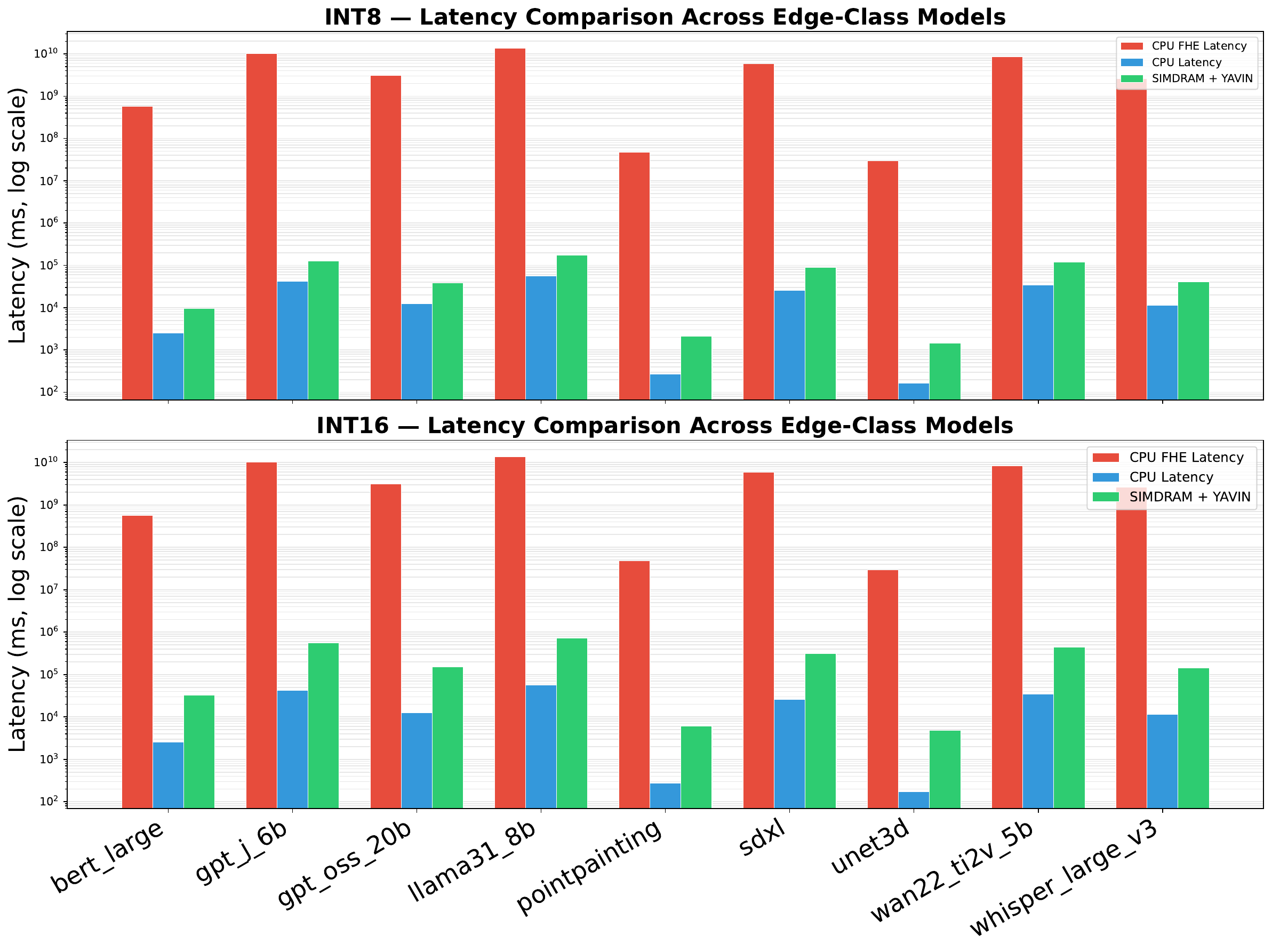}
    \caption{YAVIN Latency Comparison Across Edge-Class Models}
    \label{fig:fhe_vs_yagin}
    \vspace{-.2in}
\end{figure}

\paperName{}'s goal is to compute on protected data across untrusted memory bus in acceptable cost where fully homomorphic encryption (FHE) is the only prior approach offering comparable confidentiality in multi-tenant system, since processor-centric TEEs do not fit multiple users because of limited computing resources. Fig~\ref{fig:fhe_vs_yagin} compares three configurations across eight edge-class models at INT8 and INT16: CPU-only FHE, plaintext CPU execution, and \paperName{} (SIMDRAM bulk-bitwise computation $+$ YAVIN).

First, \paperName{} outperforms CPU FHE by \textbf{four to five orders of magnitude} (${\sim}5\times10^{4}$ speedup at INT8), turning inferences that are wholly impractical under FHE into ones that complete in seconds; by keeping plaintext only inside the trusted memory region and paying merely for lightweight symmetric encryption, \paperName{} avoids the ciphertext expansion and refresh cost that make FHE unusable on edge hardware. Second, against an \emph{unprotected} memory bus assumption, \paperName{} adds only a small constant factor ($4.1\times$ at INT8, and typically ${\sim}3\times$ for the larger models. This CPU's execution baseline is moreover for a single tenant: a CPU-based TEE serves every tenant through the same processor, so its per-tenant throughput degrades under contention, whereas \paperName{} performs both computation and crypto calculation inside the memory and scales with the number of independent banks and subarrays serving concurrent tenants. \paperName{} thus occupies an stage neither alternative reaches: FHE-like computation on protected data across an untrusted bus, at a cost within a small factor ($4 \times$) of native CPU execution and orders of magnitude below FHE, with post-quantum authenticated protection.

\subsection{Application Performance: LLM Inference}
\label{sec:eval-apps}

We evaluate \paperName{} on seven large language models spanning a range of architectures and scales, from 4B to 70B parameters: GPT-J-6B, GPT-OSS-20B, LLaMA-3.1-8B, LLaMA-3.3-70B, LLaMA-4-Scout, Mixtral-8x7B, and Qwen3-VL-4B. Each model is evaluated under quantization levels INT8 and INT32. We assume GEMM layers are computed using PIM and non-GEMM layers such as ReLU and Softmax are computed using the processor. 

\paragraph{Methodology}
PIM GEMM measures the baseline PIM inference latency without encryption under the SIMDRAM architecture. Under \paperName{}, data must be decrypted when leaving PIM for processor-side non-GEMM computation and re-encrypted upon return, at the measured ASCON throughput of 23.2\,MB/s. We define \emph{YAVIN overhead} as the ratio of total additional latency (PIM crypto + processor crypto + bit-serial/bit-parallel roundtrip) to the PIM GEMM baseline.

\begin{figure}[tbp]
    \centering
    \includegraphics[width=\linewidth]{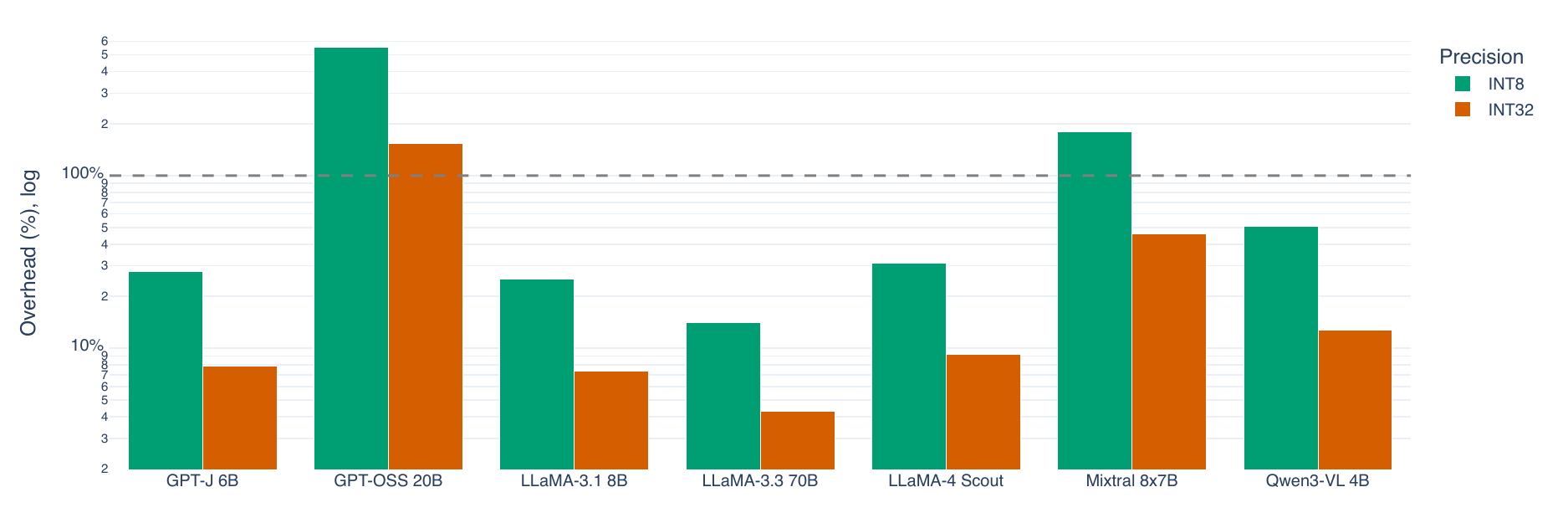}
    \caption{YAVIN's PIM encryption/decryption overhead in LLM tasks}
    \vspace{-.2in}
\end{figure}

\paragraph{Results}
PIM-side crypto dominates the overhead, accounting for $\sim$97\% of the total additional cost (Table~\ref{tab:llm-overhead}). At INT8, overhead ranges from \textbf{14\%} (LLaMA-3.3-70B) to \textbf{556\%} (GPT-OSS-20B); at INT32, it drops to \textbf{4.3\%--154\%} for the same models. The wide spread reflects each model's GEMM-to-non-GEMM ratio: models with a higher proportion of GEMM arithmetic amortize the fixed PIM encryption cost more effectively. Overhead also falls with increasing precision because compute scales faster than the encrypted parameter volume with wider operands. For instance, LLaMA-4 Scout drops from 31\% at INT8 to 9.3\% at INT32. Focusing on the edge-class workloads \paperName{} is able to provide a unified trusted computing base with charge-sharing based PIM for $34.2\%$ and $9.3\%$ overhead for 8- and 32-bit quantization, respectively. The geomean across all benchmarks remains tractable at $82\%$  and $27.7\%$ for 8- and 32-bit, respectively.

\section{Conclusion}
\label{sec:conclusion}
 
We presented \paperName{}, a unified trusted computing base that extends trusted execution beyond the processor to encompass processing-in-memory while treating the memory bus as untrusted. By combining post-quantum key establishment with lightweight authenticated encryption and co-designing both with DRAM-PIM execution, \paperName{} enables data to be decrypted, processed, and re-encrypted entirely within trusted processor or memory regions without exposing plaintext on the memory bus. Our evaluation demonstrates that these capabilities can be realized with practical overheads while preserving the performance advantages of processing-in-memory, providing a foundation for secure multi-tenant PIM systems for edge-class computing environments.



\bibliographystyle{IEEEtran}
\bibliography{refs}

\end{document}